\documentclass{sigchi}
\newcommand{\compresslist}{
  \setlength{\itemsep}{3pt}
  \setlength{\parskip}{0pt}
  \setlength{\parsep}{0pt}
}

\CopyrightYear{2018} 
\setcopyright{acmcopyright} 
\conferenceinfo{MobileHCI '18,}{September 3--6, 2018, Barcelona, Spain}
\isbn{978-1-4503-5898-9/18/09}\acmPrice{\$15.00}
\doi{https://doi.org/10.1145/3229434.3229441}

\usepackage{balance}       
\usepackage{graphics}      
\usepackage[T1]{fontenc}   
\usepackage{txfonts}
\usepackage{mathptmx}
\usepackage[pdflang={en-US},pdftex]{hyperref}
\usepackage{color}
\usepackage{booktabs}
\usepackage{textcomp}
\usepackage{epstopdf}
\usepackage{subfigure}

\usepackage{microtype}        
\usepackage{ccicons}          

\usepackage{todonotes}

\usepackage{multirow}

\usepackage{graphicx}
\usepackage{epstopdf}
\usepackage{algorithm}
\usepackage{url}
\usepackage{array}
\usepackage{algorithmicx}
\usepackage{algpseudocode}
\usepackage{paralist}
\usepackage{booktabs}

\newcolumntype{L}[1]{>{\raggedright\let\newline\\\arraybackslash\hspace{0pt}}m{#1}}
\newcolumntype{C}[1]{>{\centering\let\newline\\\arraybackslash\hspace{0pt}}m{#1}}
\newcolumntype{R}[1]{>{\raggedleft\let\newline\\\arraybackslash\hspace{0pt}}m{#1}}
\usepackage{dcolumn}

\newcolumntype{d}{D{.}{.}{3,2}}

\def\plaintitle{Typical Phone Use Habits:\\ Intense Use Does Not Predict Negative Well-Being}

\def\emptyauthor{}
\def\plainkeywords{Clustering; User groups; Mobile phone use; Well-being; Personality traits; User study; Unsupervised-learning; Machine-learning; Multi-level modelling}

\makeatletter
\def\url@leostyle{%
  \@ifundefined{selectfont}{
    \def\UrlFont{\sf}
  }{
    \def\UrlFont{\small\bf\ttfamily}
  }}
\makeatother
\def\pprw{8.5in}
\def\pprh{11in}

\definecolor{linkColor}{RGB}{6,125,233}
\hypersetup{%
  pdftitle={\plaintitle},
  pdfauthor={\emptyauthor},
  pdfkeywords={\plainkeywords},
  pdfdisplaydoctitle=true, 
  bookmarksnumbered,
  pdfstartview={FitH},
  colorlinks,
  citecolor=black,
  filecolor=black,
  linkcolor=black,
  urlcolor=linkColor,
  breaklinks=true,
  hypertexnames=false
}

\usepackage{color}
\definecolor{Orange}{rgb}{0.9,0.5,0}
\definecolor{NavyBlue}{rgb}{0.1, 0.4, 0.8}
\definecolor{Magenta}{rgb}{0.8, 0.1, 0.6}
\definecolor{Green}{rgb}{0.1, 0.8, 0.3}
\definecolor{DarkGreen}{rgb}{0.0, 0.7, 0.2}
\definecolor{Brown}{rgb}{0.4, 0.3, 0.1}
\definecolor{Burgundy}{rgb}{0.5, 0.0, 0.13}
\definecolor{BrightCerulean}{rgb}{0.11, 0.67, 0.84}

\begin{document}

\title{\plaintitle}

\numberofauthors{3}
\author{%
  \alignauthor{Kleomenis Katevas\thanks{This work was done during a student internship at Telef\'onica Research.}\\
    \affaddr{Imperial College London}\\
    \affaddr{London, UK}\\
    \email{k.katevas@imperial.ac.uk}
    }\\
  \alignauthor{Ioannis Arapakis\\
    \affaddr{Telef{\'o}nica Research}\\
    \affaddr{Barcelona, Spain}\\
    \email{ioannis.arapakis@telefonica.com}
    }\\
  \alignauthor{Martin Pielot\\
    \affaddr{Telef{\'o}nica Research}\\
    \affaddr{Barcelona, Spain}\\
    \email{martin.pielot@telefonica.com}
    }\\
}

\maketitle

\begin{abstract}

Not all smartphone owners use their device in the same way. In this work, we uncover broad, latent patterns of mobile phone use behavior. We conducted a study where, via a dedicated logging app, we collected daily mobile phone activity data from a sample of 340 participants for a period of four weeks. Through an unsupervised learning approach and a methodologically rigorous analysis, we reveal five generic phone use profiles which describe at least 10\% of the participants each: \emph{limited use}, \emph{business use}, \emph{power use}, and \emph{personality-} \& \emph{externally induced problematic use}. We provide evidence that intense mobile phone use alone does not predict negative well-being. Instead, our approach automatically revealed two groups with tendencies for lower well-being, which are characterized by nightly phone use sessions.

\end{abstract}

\category{H.5.m.}{Information Interfaces and Presentation
  (e.g. HCI)}{Miscellaneous}

\keywords{\plainkeywords}

\section{Introduction} \label{sec:intro}

The average person in the US spends approximately three hours per day on the mobile phone, while young people (18-24 years) spend significantly more\footnote{As reported at ComScore's 2017 Cross Platform Future in Focus.}. However, mobile phone overuse may lead to undesirable outcomes. Oulasvirta \emph{et al.}~\cite{Oulasvirta2012} suggested that repetitive inspection of dynamic content such as notifications from social media apps may lead to phone overuse. Other studies have reported the association of phone overuse with sleeping disorders, increased stress, depression, social isolation and decline in academic performance~\cite{Jerano:2007, Kwon:2013, Lee:2014, Lepp:2014}.

Detecting problematic phone use is a topic that has been previously examined in the research community using a variety of methods such as self-reported questionnaires~\cite{Bianchi:2005,Merlo2013}, that are known to be expensive, subjective, and prone to errors~\cite{Samkange:2004}. A different supervised learning based approach, presented by Shin and Dey~\cite{Shin:2013}, predicts these behaviours using data logs collected from the user's mobile device~\cite{Shin:2013}. This solution was based on relatively small training sample that requires an existing ground truth and is hard to generalize. In addition, it tends to treat all users as homogeneous entities, neglecting that different mobile user profiles exist.

It is important to highlight that observing an extensive mobile phone use does not directly imply that a problematic use also exists. For example, extensive phone use by a business person during work could be considered as normal behaviour compared to late night phone use of a student, a behaviour that could easily affect the student's academic performance. According to Billieux~\cite{Billieux:2012}, ``disentangling a person's mobile phone user profile is a preliminary step necessary to avoid regrouping dissimilar behaviors into a deceptive label of mobile phone addiction''.

The aim of this work is twofold: (1) uncover typical patterns of mobile phone use from a wide range of features, and (2) investigate whether clear indications of mobile phone overuse emerges spontaneously in some of these clusters.

To this end, we investigate an unsupervised approach for understanding the different types of mobile use behaviour in the daily routines of 340 participants. For an average duration of 4 weeks we collected a wide range of mobile sensor data using a custom made Android app that participants installed on their personal devices.

The main contributions of this work are:

\begin{itemize}\compresslist

\item An unsupervised learning framework for deriving natural groups of users that is rigorously validated through a wide set of metrics and statistical measures.

\item A taxonomy of five types of mobile phone users (\emph{limited use}, \emph{business use}, \emph{power use}, and \emph{personality-} \& \emph{externally induced problematic use}) that provides rich and explanatory insights with respect to mobile phone use patterns, user demographics (per class), and personality traits such as the emotional user's state (Bad/Good, Tense/Calm, Tired/Awake, Bored), Big Five Personality Test (Big5), Personal Health Questionnaire Depression Scale (PHQ-8) and Boredom Susceptibility Scale (SSS-BS).

\item Evidence that there are specific types of phone use that associated with negative well-being, such as nightly phone use sessions. 
In contrast, overall usage intensity did not predict negative well-being.

\end{itemize}

In this work, we consider a fairly large group of users (contrary to other studies) that is not bounded by a specific age range or profession, \emph{i.e.}, our sample has some diverse characteristics and thus we believe it achieves a higher degree of external validity, compared to other studies that employed smaller samples of participants. To our knowledge, we are the first to derive generic mobile phone user profiles and investigate their interaction with behavioural and mental health measures.

\section{Background and Related Work} \label{sec:related_word}

\subsection{Definition of Problematic Phone Use}

In this work, we use the term \emph{problematic phone use} to refer to the family of behaviors with undesirable effects on emotional well-being. 
Previous work has used a wide range of other terms, including problematic cellular phone use~\cite{Yen:2009}, dysfunctional use~\cite{Billieux:2015}, maladaptive use~\cite{Beranuy:2009}, and compulsive use \& technostress~\cite{Lee:2014}.
Some researchers even coined the term \emph{smartphone addiction}~\cite{Salehan:2013, Roberts:2015, Hong:2012, Billieux:2015}. For example, Park and Lee~\cite{Park:2011} argued that mobile phone overuse has similar characteristics as impulse control disorders (ICD). Problematic phone use could be considered as one form of technological addictions. Since 2013, the Diagnostic and Statistical Manual of Mental Disorders, Fifth Edition (DSM-5)\footnote{http://www.dsm5.org/} recognizes behavioural, non-substance related addictions and recommends further research into existing technological addictions for later inclusion. Kwon \emph{et al.}~\cite{Kwon:2013} give examples of smartphone addiction, such as not paying attention in traffic or not being able to concentrate in class. However, no clear uncontested definition exists for this term. With \emph{problematic phone use}, we refer to any type of mobile phone use that negatively affects its user's emotional well-being.

\subsection{Undesirable Side-Effects of Mobile Phone Use}
Previous work reported a wide range of undesirable side-effects of problematic mobile phone use on emotional well-being, \emph{e.g.}, causing psychological distress~\cite{Beranuy:2009}, depression~\cite{Yen:2009, Thomee:2011}, anxiety~\cite{Jerano:2007, Lepp:2014}, sleep disorders~\cite{Jerano:2007, Thomee:2011}, and reduced satisfaction with life~\cite{Lepp:2014}.
Beranuy \emph{et al.}~\cite{Beranuy:2009} reported an association between heavy mobile phone use and psychological distress in 404 undergraduate college students.
Jerano \emph{et al.}~\cite{Jerano:2007} examined the effects of internet and cell phone use in 337 college students studying in Spain. They reported an association of heavy Internet use with high anxiety, and high cell-phone use with high anxiety and insomnia. 
Lee \emph{et al.}~\cite{Lee:2014} found a correlation of compulsive smartphone use and psychological traits such as locus of control, social interaction anxiety, materialism and the need for touch. 
Lepp \emph{et al.}~\cite{Lepp:2014} found a negative correlation of mobile phone use and texting with academic performance, and a positive correlation to anxiety.
Yen \emph{et al.}~\cite{Yen:2009} examined the association between problematic mobile phone use and depression in a large number of 10,191 adolescent students in Taiwan. They reported that adolescents who had significant depression were more likely to have four or more symptoms of problematic mobile phone use.
In a questionnaire-based study with 300 university students, Samaha and Hawi~\cite{Samaha:2016} found correlation between smartphone addiction and stress, a negative correlation with academic performance and a mediated negative correlation with satisfaction with life.

\subsection{Predictors of Problematic Phone Use}

Socio-demographic factors have been found to predict problematic mobile phone use: female mobile phone users are more likely to develop dependence-related symptoms with regards to mobile phones~\cite{Billieaux:2008, Beranuy:2009, Lee:2014, Jerano:2007}. 
Furthermore, younger mobile phone users tend to be more likely to exhibit symptoms of dependency to mobile phone use~\cite{Bianchi:2005, Yen:2009}.

Personality traits can also be predictors of problematic phone use. In particular, problematic phone use has been correlated to neuroticism~\cite{Ehrenberg:2008}, anxiety~\cite{Jerano:2007, Hong:2012}, distress~\cite{Beranuy:2009}, and social extraversion~\cite{Hong:2012}. Billieaux \emph{et al.}~\cite{Billieaux:2008} reported that impulsivity is related to the number and duration of the calls made daily, while urgency appeared to be the strongest predictor of problematic use. Low self-esteem has also been linked with dysfunctional use~\cite{Bianchi:2005, Billieux:2012}.

Further, two studies found that patterns of phone use themselves can predict whether participants would score above a threshold on questionnaires related to undesirable emotional well-being.
Shin and Dey~\cite{Shin:2013} found that the number of apps used per day, the ratio of SMSs to calls, the number of event-initiated sessions, the number of apps used per event initiated session, and the length of non-event-initiated sessions are the best predictors for scoring high on a questionnaire about behavioral addiction. 
Lee \emph{et al.}~\cite{Lee:2014} found that \emph{e.g.}, using the phone a lot, longer period of phone use during mornings and evenings, time spent browsing the web, and ``highly skewed usage pattern with respect to a few frequently used apps'' predict whether phone users will fall into the high-risk group of the \emph{Smartphone Addiction Proneness Scale for Adults}~\cite{Kim:2012}.

Both mentioned works group people into a ``normal'' and a ``problematic'' group depending on the questionnaire based scores. This approach ensures the presence of an existing group of problematic users. Thus, the used machine-learning algorithm will do its best to identify those groups. However, in reality, problematic phone users might not stand out too much from normal users. Therefore, it remains unclear whether problematic mobile phone use is a fringe phenomenon, or whether it has manifested itself in common patterns of phone use.

\subsection{Types of Mobile Phone Use}

In order to put problematic phone use into context, it is important to know the typical mobile phone use behavior. Typical patterns of phone use have widely been reported in recent, related work. For example, Andrews \emph{et al.}~\cite{Andrews:2015} reported the mobile use patterns of 19 staff members of the University of Lincoln for a period of 2 weeks, by examining the screen on/off events collected using an Android app. They found that their participants used the phone 84.68 times each day (SD = 55.23) and spent 5.05 hours each day on it. On the basis of logs from 29,279 devices. 
Hintze \emph{et al.}~\cite{Hintze:2017} observed that the average mobile phone user has 60 sessions per day and uses the device for about 1.5 hours per day. Mathur \emph{et al.}~\cite{Mathur:2017} analyzed four broad facets of smartphone behavior in India. They found that compared to Western users, Indian users exhibit more nightly use and charge their phones more frequently.

However, not all people use their mobile phone the same way. Such general statistics as reported by the work above may not do justice to all types of mobile phone users. Thus, previous work has attempted to identify clusters of mobile phone use behavior through unsupervised-learning approaches. The rationale was that these approaches would reveal the naturally-existing, most dominant types of phone use.

Do \emph{et al.}~\cite{Do:2010} explored typical patterns of phone use. They model patterns of phone use by representing app launches by a bag-of-application model, this is, modelling how frequently it would be observed to launch different types of apps (voice, SMS, internet, camera, gallery) during four time frames: night, morning, afternoon and evening. While they identify common topics of phone use, they do not disclose common types of users.

Zhao \emph{et al.}~\cite{Zhao:2016} examined the app use behaviours in a large scale study with 106,762 Android users. Using an unsupervised learning approach, they distinguished 382 types of users based on their app usage traces (\emph{i.e.} weights of how much each app category has been used per user). They profiled the 3 biggest clusters ranked by size, and named \emph{Night communicators}, \emph{Screen checkers}, and \emph{Evening learners} -- showing that distinct patterns of usage exist. This work was limited by the absence of more broad features, such as app usage duration, phone calling interaction, or generic phone usage, information that could lead into a more precise clustering of user behaviour.

Visuri \emph{et al.}~\cite{Visuri:2017} analyzed patterns of phone use to improve the scheduling of quantified-self self-report questionnaires. On the basis of pre-study questionnaire and the device use of 48 participants, they identified five clusters: 
(a) the casual user who was most passive and typically female; 
(b) the social chatterer who was rather young and frequently used communication apps; 
(c) a balanced cluster with no clear characteristics; 
(d) the night owl who tended to use the phone during the night and frequently used communication apps; and 
(e) the work on-the-go user who was an active phone user that quickly responded to interruptions. Clusters 6 and 7 were discarded, because they had no distinct features and only few users. While not their primary outcome, the results of Visuri \emph{et al.} indicate that mobile phone users can be separated into a handful of meaningful clusters.

None of these previous works investigated to what extent these types of behaviors are associated with problematic phone use and undesirable effects on emotional well-being.

\section{Experimental Methodology}
\label{sec:experimental_methodology}

The first goal of this work is to identify clusters of \emph{typical} patterns of phone use. The objective is to keep our natural groupings of users as high-level and general as possible so that they can be easy for interpretation and further analysis. Treating every person as individual exceeds our capacity to effectively design better-adapted solutions. Hence, we seek to identify a few, broad clusters that cover the most dominant patterns of mobile phone use. To this end, we introduced the constraint that clusters of typical phone use should at least contain 10\% of the study population.

The second goal is to explore whether problematic phone use spontaneously emerges as a cluster. If problematic phone use is a real and wide-spread problem which is clearly related to how the mobile phone is used, we would expect that it clearly stands out as a cluster as well. This would provide evidence for the existence of problematic phone use as a relevant problem to the society. To this end, we apply an unsupervised-learning pipeline onto a data set of mobile phone use logs.

\subsection{Participants}
\label{ssec:participants}

Our data set contains mobile phone use logs from 340 Android phone users, recruited through a specialized agency for an average duration of four weeks during summer 2016. We requested a sample that matches the gender and age distribution of Spain, the country of study. The only restriction was that people were required to own an Android phone as Android phones account for the large majority ($\sim$90\%) of the smartphone share in the country of study. The participants' ages ranged from 18 to 66 years ($M = 37.850$, $SD = 11.025$), with a balanced gender split (53.24\% female and 46.76\% male).

\subsection{Data}
\label{ssec:data}

The data was collected through an app which was running in the background while passively collecting rich sensor data about the user's context and phone use. Participants registered the app to listen for accessibility events, which allowed it to log which apps they were using. In addition, the app collected information about the user's phone use, including the screen status (on, off, unlocked), phone calls (incoming, outgoing), data activity (kb/sec), ringer mode (default, vibration, silent), battery level and number of photos taken. Basic demographics such as age and gender were self-reported at the beginning of the study and included in the data set.

\subsection{Experience sampling}
\label{ssec:esm}

In addition to sensor-data logging, the application collected self-report via experience sampling. The application created about 10 to 15 notifications per day. The notification text prompted users to report their emotional state by asking ``How do you feel?''. 

Clicking on such a notification opened a questionnaire with 4 Likert-scale items about the emotional state of the user. More specifically, the user reported whether they are feeling: Bad/Good, Tense/Calm, and Tired/Awake, using a five-point scale (reduced into a three-point scale, \emph{i.e.} Bad/Neutral/Good), as well as feeling Bored using a boolean scale (No/Yes).

The timing of the delivery was semi random to capture a wide range of scenarios. Between 22:00 and 8:00 in participants' local time, the notifications were silent and without vibration. After 6 responses of a given day, notifications were suspended and resumed only on the next day. If the participants did not respond within 10 minutes, the notification was removed from the notifications tray. To be eligible for the compensation, participants were required to respond to at least two of these questionnaires for a minimum of 21 days.

In addition to asking participants to report their mood, on the bottom of the questionnaires, the users were offered additional things to do. Amongst them was an option to ``learn more about yourself'' that led to psychometric questionnaires sequentially selected from the following: Big Five Personality Test (Big5), Personal Health Questionnaire Depression Scale (PHQ-8) and Boredom Susceptibility Scale (SSS-BS).
The Big5~\cite{BIG5} scale consists of 20 questions that evaluates the user's personality in five domains: \emph{Openness} (to experience), \emph{Conscientiousness}, \emph{Extraversion}, \emph{Agreeableness} and \emph{Neurocism}, using a scale from 0 (low) to 24 (high).
PHQ-8~\cite{PHQ8} is an 8-item questionnaire that is used to detect the presence and severity of depression in patients, according to DSM-5. The scale ranges between 0 (low likelihood) and 24 (high likelihood).
SSS-BS~\cite{zuckerman1979sensation} measures aversion to routine, repetition and dull people, using a scale that ranges between 0 (low boredom propensity) and 66 (high boredom propensity). Responding to these questionnaires was voluntary. Hence, we only collected responses for a subset of the participants.

The case study has been approved by the legal department of our institution. Taking user security and privacy seriously, all communication with our back-end service was secured using SSL encryption (TLS v1.2) and all data collected was stored in our database using pseudonymization, being impossible to identify and reach out participants.

\subsection{Features}
\label{ssec:features}

Phone use directly reflects in the number of phone unlocks, app launches, calls received etc. Since our focus was to investigate typical phone use, we computed the daily median of 23 of such features (\emph{e.g.} daily number of phone sessions, median duration of calls, battery drain etc.), as well as three features that characterize the device status during the data collection period (\emph{i.e.} status of the ringer mode). We chose daily features over hourly or weekly as we believed that hourly features would be too fine-grained and weekly too coarse-grained information. Moreover, we chose the median as a measure of central tendency since most of the scores of the features are not normally distributed.

One defining property of overuse and addictive behavior is that it leads us to override and ignore our basic physiological needs, such as sleep. This, in turn, leads to negative outcomes, such as reduced well-being~\cite{Fossum:2014}. Thus, we separately investigate phone use that took place during night time (\emph{i.e.} 0:00 -- 5:59) as a proxy for interactions that were likely to have interrupted sleep. All interactions happening between 6:00 and 23:59 will be referred to as ``day'' features, interactions between 0:00 and 5:59 as ``night'' features for the remainder of this work. We ended up with 26 features in total (see Table~\ref{tbl:descriptive_statistics}) that belong to the following categories of mobile sensor events:

\textit{Screen sessions} -- 4 features capturing the number of times a user unlocked the screen, as well as the duration of phone activity, during the day and night separately. These features indicate the general amount and frequency of phone use.

\textit{Phone calls} -- 8 features representing the number and duration of outgoing and incoming calls as an indication of perceived dependence on the phone~\cite{Billieaux:2008}.

\textit{App launches} -- 8 features modeling the number of app launches per day \& night of 4 categories: \emph{Social} -- as frequently checking social media apps has been previously associated with negative outcomes~\cite{Oulasvirta2012}, \emph{Messaging} -- as an indication of personal and direct communication, \emph{Email} -- as an indicator of professional communication, and \emph{Games} -- as an indication of time killing in recreation activities. We excluded other type of apps as general phone use can still be captured by the `screen sessions' feature listed above.

\textit{Network data activity} -- captures the average daily download volume (in kb/sec), serving as a proxy to the degree of phone use intensity. Previous works have shown that data activity was an important predictor of negative effects in well-being such as boredom~\cite{Pielot:2015}.

\textit{Photos taken} -- models the daily number of pictures taken with the phone. Extensive use of the phone's camera can indicate that users are occupied with interesting, picture-worthy activities and/or that they are communicating activities to friends via photo messages or social media posts. Moreover, the popular selfie (self-portrait photograph) trend has been recently associated with narcissism, self-perception of leadership, vanity and exhibitionistic propensities~\cite{weiser2015me}. Note that only the quantity of pictures taken per day was accessed, without looking at the content of the pictures itself.

\textit{Battery drain} -- represents the percentage of the device's battery drain levels per hour. High battery drain indicates heavier use of the phone (\emph{e.g.} engagement with games, communication using video calling apps etc.).

\textit{Ringer mode} -- a one-hot-encoded feature representing the fraction of time that a device was set to normal, silent and vibrate ringer mode during the study. These features are an indication of how tolerant a user is to disruptions.

\vspace{-0.01cm}
\subsection{Unsupervised Learning}
\label{ssec:unsupervised_learning}
The main objective of this analysis is to divide the users in our data set into natural groups that reflect salient patterns of mobile phone use. To this end, we employed unsupervised learning to derive generic profiles of mobile phone users, \emph{i.e.}, profiles that abstract from characteristics of an individual to behaviours and habits shared by the members of each group. 

Given that there are no predefined classes, the outcome of a clustering process is not (always) deterministic and may result in a different partitioning of the data, depending on the specific criteria used each time (\emph{e.g.}, number of desirable clusters or type of clustering algorithm). Since we do not know \emph{a-priori} which clustering algorithm and configuration settings will perform better for the examined domain, we opted for several representative options. The reason was to avoid the common pitfall of resorting to a single clustering algorithm that may not be able to produce proper partitions, due to the nature of the given data set.

To understand which clustering algorithm performs best, we evaluated the result produced by each clustering algorithm using several internal validity metrics.
Since each validity metric can result in a different ranking of the clustering configurations, we reached a consensus for the best clustering setup through rank aggregation. This gave us a super list which was as ``close'' as possible to all individual ordered lists simultaneously, with the top positions granted to the clustering configurations that performed better amongst many validity metrics.

\begin{table*}[t!]
    \centering
    \caption{Clustering Algorithm Parameters.}
    \small
    \begin{tabular}{| l | l | l | l |}
    \hline
    \bf Category        & \bf Algorithm        & \bf Parameters      & \bf Value  \\
    \hline \hline
    Centroid              & {\tt K-means}              & \#~Clusters        & [2, 5]     \\ \hline 
    \multirow{5}{*}{Connectivity}
                         & \multirow{3}{*}{{\tt Agglomerative}} & \#~Clusters        & [2, 5]     \\
                         & & Linkage type       & \it ward, complete, average \\
                         & & Affinity           & \it euclidean, l1, l2, manhattan, cosine \\ \cline{2-4}
                         & \multirow{2}{*}{{\tt Spectral}} & \#~Clusters        & [2, 5]     \\
                         & & Kernel type        & \it nearest neighbors, rbf \\ \hline

    \multirow{3}{*}{Density}
                         & \multirow{2}{*}{{\tt DBSCAN}} & Neighborhood size  & \it 1e-6, 1e-5, 1e-4, 5e-3, 1e-3, 4e-2, 3e-2, 2e-3, 1e-4           \\
                         & & Min samples        & 34 (10\% of the user set)               \\ \cline{2-4}
                         & {\tt Mean-shift}           & Bandwidth          & Auto using \textit{0.1, 0.25, 0.5, 0.75, 0.9} quantiles  \\ \hline
    \multirow{3}{*}{Distribution}    
                         & \multirow{3}{*}{{\tt Gaussian Mixtures}} & \#~Components      & [2, 5]     \\
                         & & Covariance type    & \it full, tied, diagonal, spherical \\
                         & & Convergence threshold  & \it 1e-4, 1e-3, 1e-2, 1e-1 \\ \hline
    \end{tabular}
    \label{tab:clustering_algorithms}
    \vspace{-0.2cm}
\end{table*}

\subsubsection{Preprocessing}
\label{sssec:preprocessing}
For the purposes of the unsupervised learning task, we aggregated our data per user and per day, and computed the median values for all extracted features. Given that some of the non-agglomerative clustering algorithms that we considered (Table~\ref{tab:clustering_algorithms}) work by optimizing a distance criterion (\emph{e.g.}, Euclidean distance), we normalized our features, which homogenized their relative importance in the clustering experiments.

\subsubsection{Clustering Algorithms}
\label{sssec:clustering_algorithms}

Table~\ref{tab:clustering_algorithms} summarizes the clustering algorithms and the ranges of values we used to parametrize them. More specifically, we considered centroid-, connectivity-, density-, and distribution-based clustering algorithms. \emph{Centroid-based} clustering algorithms derive the notion of similarity by the closeness of a data point to the centroid of the clusters. These algorithms run iteratively to find the local optima and the the number of clusters required in advance. \emph{Hierarchical-based} (aka \emph{connectivity-based}) clustering algorithms work by forming an initial pair of clusters and then recursively consider whether it is worth splitting each one further; this type of clustering algorithms produce a hierarchy that can be represented as a binary tree (\emph{i.e.}, dendrogram). \emph{Density-based} clustering algorithms search the data space for areas of varied density of data points in the data space. They isolate various different density regions and assign the data points within these regions in the same cluster. Last, the \emph{distribution-based} clustering algorithms do not place data instances categorically in one cluster or the other; instead, they assign them a certain probability of belonging to each cluster.

The set of clustering algorithms $\times$ their parameter settings (see Table~\ref{tab:clustering_algorithms}) results in 135 possible clustering configurations. From this set, we excluded those clustering configurations that produced clusters with less than 10\% of the users, namely {\tt Mean-shift} and {\tt DBSCAN}. These clustering algorithms almost always resulted in as many clusters as there are data points and, thus, were unfit for our analysis. 
In what follows, we compute several cluster-validity criteria to assess the quality of the clustering and establish whether the identified groups reflect typical and distinguishable types of phone use.

\subsubsection{Measuring Cluster Validity}
\label{ssec:measuring_cluster_validity}
Due to the unsupervised nature of clustering methods, results may vary greatly in terms of the data partitioning. Therefore, an important challenge in cluster analysis is the evaluation of the quality of the results. The ``correctness'' of clustering algorithm results is verified using a process known as cluster validation~\cite{Halkidi2001}. Three main approaches exist for this purpose: (1) external criteria, (2) internal criteria, and (3) relative criteria. In this work, we evaluated the results of our clustering experiments using the second type of validity criteria (internal), which consider only quantities and features inherent to the data.  This choice was based on the larger availability of clustering validity criteria that do not rely on an external ground truth.

\begin{table}[t!]
    \centering
    {\small
    \caption{Internal validity criteria used for evaluating the quality of the produced clusters}
    \vspace{0.1cm}
    \label{tbl:internal_validity_criteria}
    \begin{tabular}{@{} L{24mm} C{12mm} L{24mm} C{12mm} @{}}
    \toprule
    \textbf{Algorithm} & \textbf{Rule} & \textbf{Algorithm} & \textbf{Rule} \\
    \midrule
    Banfeld Raftery & \textit{min} & Point Biserial &  \textit{max}   \\
    C index & \textit{min} & Davies-Bouldin & \textit{min}    \\
    Dunn & \textit{max} & Ray-Turi & \textit{min}    \\
    Gamma & \textit{max} & SD-Scat & \textit{min}    \\
    log(BGSS/WGSS) & \textit{min diff} & Silhouette & \textit{max}   \\
    McClain-Rao & \textit{min} & Xie-Beni & \textit{min}    \\
    PBM & \textit{max} &  &    \\
    \bottomrule
    \end{tabular}
    }
    \vspace{-0.2cm}
\end{table}

If multiple internal validity criteria are being used, there is a risk that some of these optimize a similar cost function. This can be a problem if the type of cost function is not suitable for the data set: if criteria with sub-optimal cost functions dominate the set of evaluation criteria, they will bias the overall evaluation and lead to a poor choice of the final clustering. To mitigate this, we removed the subset of criteria that optimized very similar cost functions or produced highly-correlated rankings. This is similar to the ensemble learning approach, where instead of relying on one machine-learning algorithm to make decisions, we would combine the output of several different models by learning an ensemble of models and using them in combination. This process resulted in to a selection of 13 widely-used criteria that are shown in Table~\ref{tbl:internal_validity_criteria}.

\subsubsection{Selecting Validity Measures}
\label{sssec:selecting_validity_measures}

In previous section, we addressed how we measure the quality of the produced clustering configurations. Prior to applying the rank aggregation algorithm, we downsized our internal validity metrics to a subset that shows low consistency among the produced rankings. This is because (1) using all available internal validity metrics can slow down rank aggregation and (2) including highly correlated internal validity metrics can bias the results towards a specific type of cost function and eventually affect their quality. To this end, we computed all $\frac{n(n-1)}{2}$ pairwise Spearman correlations of the orderings. This gave a complete weighted graph $G=(V,E)$: vertices are the validity measures and each edge $e = (v_i, v_j)$ is weighted with the correlation value between $v_i$ and $v_j$.
Given $G$, we selected the $k$ vertices, such that the resulting $k$-clique has the minimum edge-weight. The minimum edge-weight clique problem is NP-Hard~\cite{Ravi1994}. Therefore, we used the greedy approximation shown in Algorithm~\ref{alg:sel-val-meas}. The output contained the $k$ validity measures with the lowest overall agreement.

\subsubsection{Rank Aggregation}
\label{sssec:rank_aggregation}

Given the $k$ validity measures with high consistency, we wanted to identify the $n$ clustering configurations that were better (\emph{i.e.}, were ranked higher) than the rest. More specifically, we needed those clustering configurations that had a better standing in as many lists as possible. This is a typical instance of \textit{multiple-winner voting problem}, where we are given a set of ordered preferences as input (ordered clustering configurations per measure in our case), and we want to select $n$ winners. For our analysis, we chose rank aggregation~\cite{Dwork2001}, which has ties to other known techniques (e.g., \textit{Borda Count}, \textit{Condorcet}, etc.), but is better at filtering out noise. Furthermore, we used the R implementation \emph{RankAggreg}~\cite{RankAggreg} with the settings recommended by~\cite{Pihur2007}, which is an efficient implementation of the algorithm despite of the problem's NP-Hardness.

We can formalize our goal within the framework of the following optimization problem: We want to find a super list which would be as ``close'' as possible to all individual ordered lists simultaneously.

The objective function that satisfies it is the following:
\begin{equation}
\label{equation:objective_function}
\Phi(\delta) = \sum_{i=1}^{m}{w_{i}d(\delta, L_{i})}
\end{equation}
\vspace{-0.3cm}

where $\delta$ is a proposed ordered list of length $k = |L_{i}|$, $w_{i}$ is the importance weight associated with list $L_{i}$, $d$ is a distance function, and $L_{i}$ is the $i^{th}$ ordered list. The idea is to find $\delta^{*}$ which would minimize the total distance between $\delta^{*}$ and $L_{i}$'s:

\vspace{-0.3cm}
\begin{equation}
\label{equation:rankaggregation}
\delta^{*} = \arg min \sum_{i=1}^{m}{w_{i}d(\delta, L_{i})}
\end{equation}
\vspace{-0.3cm}

In our case, the items of the lists are the clustering configurations and each ranked list $L_{i}$ is a internal validity metric. We ranked the lists in ascending order according to the score that each clustering configuration achieved, and subsequently produced their aggregated ranking using the Cross-Entropy Monte Carlo algorithm~\cite{Boer:2005aa,Rubinstein2004}. The Cross-Entropy Monte Carlo algorithm is an iterative procedure for solving difficult combinatorial problems in which it is computationally not feasible to find the solution directly. We applied a non-weighted aggregation that optimized a distance criterion, \emph{e.g.} Kendall's tau and Spearman's Footrule distance, and allowed for a far more objective and automated assessment of the clustering results. The result was a list with the top positions granted to the clustering configurations that performed better than others, for the internal validity metrics that we considered. The top-ranked clustering configuration was the {\tt Spectral} with rbf kernel, for $k = 5$, which we used in the subsequent analysis.

\begin{algorithm}[t]
{\small
  \caption{{\tt \scriptsize{Greedy approximation algorithm for selecting the internal validity measures with lowest agreement.}}}
  \label{alg:sel-val-meas}
  \begin{algorithmic}[1]
    \Statex {\textbf{Input:} Weighted Graph $G=(V,E)$, int $k$ }
    \Statex {\textbf{Output:} Max-Edge Weight Subgraph $G_k$, $|G_k| = k$ }

	\State $G_k \leftarrow \emptyset$
	\State $e_1 \leftarrow$ Select edge with maximum weight.
	\State Add vertices of $e_1$ to $G_k$
	\While {( $|G_k| < k $ )} 
		\For { $v_i \in (G \ G_k) $ }
			\State $s_i \leftarrow \sum_{v_j \in G_k} weight(v_i, v_j)$
		\EndFor
		\State Select $v_j$ with $\max s_i$
		\State $G_k \leftarrow (G_k \cup v_j)$
	\EndWhile
	\State \textbf{return} $G_k$
 \end{algorithmic}
}
\end{algorithm}

\subsection{Multilevel Modelling}
\label{ssec:multilevel_modelling}
Research studies that span across longer periods of time require a different way of thinking about design and analysis (\emph{e.g.}, equating time as an independent variable, accounting for correlated errors) than do cross-sectional designs, even to the extent that the data sets must be structured differently. To this end, we analyze the mood and psychometric questionnaire data we collected throughout our study by means of mixed multilevel models, a method that allows dealing with some specificities of repeated measures data (violation of sphericity, nested data, autocorrelation of residuals) more effectively compared to traditional statistical methods such as ANOVA~\cite{kristjansson2007}. Mixed models involve looking at how individuals (or units, groups, organizations, etc.) change over time and whether there are differences in patterns of change.
 
For our analysis, we used the R package \emph{nlme}~\cite{nlme}. Our approach to the model construction was the one described in~\cite{bliese2002, bliese2013, kristjansson2007} by fitting a series of two-level models, in which measurements for each condition (level 1) were nested within participants (level 2). Initially, for our level 1 analysis, we began with a random intercept model for all of our predictors (ESM, PHQ-8, etc.) and estimated the intra-class correlation coefficient (ICC) to determine the strength of the non-independence of the observations. The unconditional means model provides between-group and within-group variance estimates in the form of $\tau_{00}$ and $\sigma^{2}$, respectively. To determine whether $\tau_{00}$ is significant, we compared -2 log likelihood values between (1) a model with a random intercept, and (2) a model without one.

\begin{table*}[t!]
    \centering
    {\small
    \caption{Descriptive statistics (Mean, Std.) for the produced clusters of mobile phone users}
    \label{tbl:descriptive_statistics}
    \begin{tabular}{@{} L{4.8cm} C{1.8cm} C{1.8cm} C{1.8cm} C{1.8cm} C{1.8cm} C{1.8cm} @{}}
    \toprule
    \textbf{Feature} & \textbf{C1} & \textbf{C2} & \textbf{C3} & \textbf{C4} & \textbf{C5} & \textbf{All} \\
    \midrule
    $f_{1}$: App Launched Day: \# Social & 3.64~(6.65) & 4.54~(7.29) & 16.34~(31.17) & 9.40~(12.86) & 9.33~(14.06) & 8.43~(17.28) \\
    $f_{2}$: App Launched Day: \# Messaging & 17.85~(15.88) & 45.41~(33.27) & 44.83~(55.30) & 40.35~(30.75) & 39.20~(29.78) & 35.12~(35.82) \\
    $f_{3}$: App Launched Day: \# Email & 1.83~(3.25) & 4.72~(6.78) & 8.52~(10.38) & 2.99~(3.86) & 4.47~(9.35) & 4.26~(7.44) \\
    $f_{4}$: App Launched Day: \# Games & 0.49~(2.07) & 0.59~(1.56) & 4.40~(7.97) & 1.10~(2.82) & 2.43~(6.35) & 1.77~(5.04) \\
    $f_{5}$: App Launched Night: \# Social & 0.11~(0.56) & 0.47~(1.31) & 2.19~(6.96) & 1.27~(3.34) & 0.80~(2.14) & 0.91~(3.63) \\
    $f_{6}$: App Launched Night: \# Messaging & 0.22~(0.63) & 0.59~(1.61) & 2.90~(9.60) & 3.37~(7.30) & 2.11~(4.43) & 1.74~(5.78) \\
    $f_{7}$: App Launched Night: \# Email & 0.00~(0.00) & 0.00~(0.00) & 0.33~(0.64) & 0.02~(0.13) & 0.00~(0.00) & 0.07~(0.31) \\
    $f_{8}$: App Launched Night: \# Games & 0.01~(0.10) & 0.10~(0.36) & 0.22~(0.64) & 0.15~(0.81) & 0.13~(0.39) & 0.11~(0.50) \\
    \midrule
    $f_{9}$: Battery Drain (\% per hour) & 6.14~(5.03) & 9.63~(5.60) & 18.16~(11.28) & 12.01~(6.61) & 11.54~(6.21) & 11.11~(8.31) \\
    \midrule
    $f_{10}$: Calls Day: \# Incoming & 0.66~(0.85) & 4.34~(2.41) & 0.91~(0.92) & 0.67~(0.87) & 0.94~(1.54) & 1.25~(1.79) \\
    $f_{11}$: Calls Days: Incoming Duration (min) & 0.60~(0.95) & 6.85~(5.57) & 0.88~(1.59) & 0.43~(0.79) & 1.05~(3.06) & 1.54~(3.33) \\
    $f_{12}$: Calls Day: \# Outgoing & 0.33~(0.61) & 4.46~(3.91) & 0.48~(0.68) & 0.51~(1.01) & 0.82~(1.43) & 1.03~(2.14) \\
    $f_{13}$: Calls Day: Outgoing Duration (min) & 0.37~(1.02) & 8.52~(7.11) & 0.51~(1.07) & 0.75~(1.94) & 1.06~(2.03) & 1.68~(3.97) \\
    $f_{14}$: Calls Night: \# Incoming & 0.00~(0.00) & 0.00~(0.00) & 0.00~(0.00) & 0.00~(0.00) & 0.01~(0.06) & 0.00~(0.03) \\
    $f_{15}$: Calls Night: Incoming Duration (min) & 0.00~(0.00) & 0.00~(0.00) & 0.00~(0.00) & 0.00~(0.00) & 0.00~(0.03) & 0.00~(0.01) \\
    $f_{16}$: Calls Night: \# Outgoing & 0.00~(0.00) & 0.00~(0.00) & 0.00~(0.00) & 0.00~(0.00) & 0.00~(0.00) & 0.00~(0.00) \\
    $f_{17}$: Calls Night: Outgoing Duration (min) & 0.00~(0.00) & 0.00~(0.00) & 0.00~(0.00) & 0.00~(0.00) & 0.00~(0.00) & 0.00~(0.00) \\
    \midrule
    $f_{18}$: Network Activity: Received (kB/s) & 0.13~(0.27) & 0.10~(0.16) & 0.29~(0.42) & 0.11~(0.16) & 0.11~(0.26) & 0.15~(0.28) \\
    \midrule
    $f_{19}$: Photos: \# Taken & 0.65~(1.19) & 0.74~(0.99) & 1.47~(2.25) & 0.85~(1.47) & 0.81~(1.25) & 0.89~(1.52) \\
    \midrule
    $f_{20}$: Ringer Mode: \% Normal & 0.93~(0.10) & 0.85~(0.20) & 0.77~(0.22) & 0.29~(0.22) & 0.22~(0.19) & 0.63~(0.35) \\
    $f_{21}$: Ringer Mode: \% Silent & 0.02~(0.05) & 0.06~(0.11) & 0.07~(0.11) & 0.55~(0.23) & 0.07~(0.11) & 0.14~(0.23) \\
    $f_{22}$: Ringer Mode: \% Vibrate & 0.05~(0.08) & 0.09~(0.16) & 0.17~(0.18) & 0.14~(0.15) & 0.71~(0.18) & 0.23~(0.29) \\
    \midrule
    $f_{23}$: Sessions Day: \# Sessions & 22.36~(14.42) & 43.67~(25.09) & 47.90~(29.67) & 37.11~(22.43) & 39.26~(24.08) & 36.23~(24.68) \\
    $f_{24}$: Sessions Day: Duration (min) & 73.45~(57.06) & 177.74~(152.01) & 293.06~(224.24) & 162.85~(146.01) & 164.43~(118.75) & 164.76~(161.32) \\
    $f_{25}$: Sessions Night: \# Sessions & 0.97~(1.27) & 1.24~(1.57) & 3.10~(3.92) & 2.69~(4.68) & 2.33~(3.97) & 2.01~(3.39) \\
    $f_{26}$: Sessions Night: Duration (min) & 2.47~(4.70) & 12.75~(39.51) & 63.68~(129.82) & 26.56~(81.57) & 22.34~(68.50) & 24.20~(77.89) \\
    \bottomrule
    \end{tabular}
    }
\end{table*}

\begin{figure*}[t!]
\centering
\begin{tabular}{c}
\hbox{
\includegraphics[trim=0 0 0 0, clip, width=0.27 \linewidth]{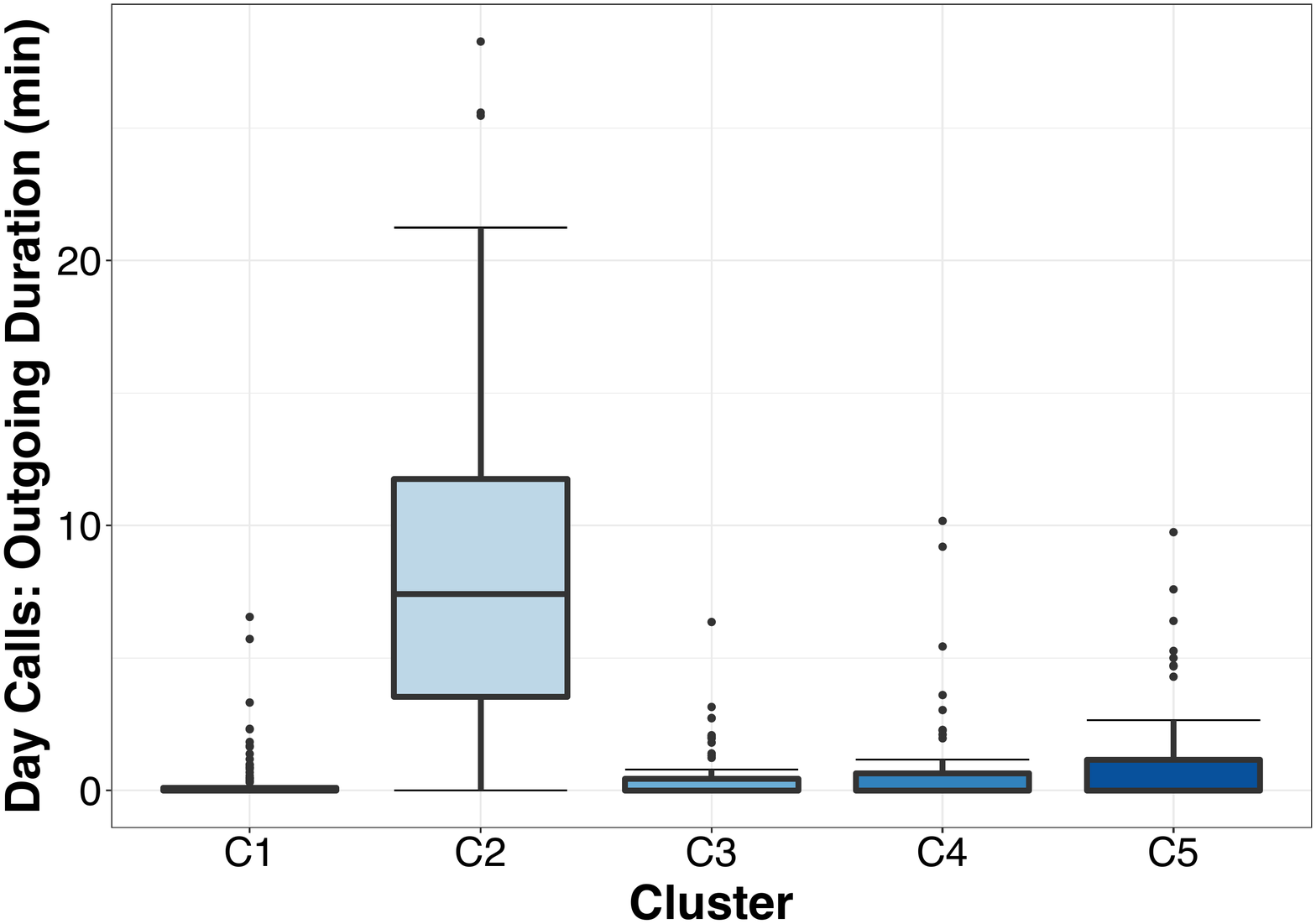}\label{feature-1}
\hfil
\includegraphics[trim=0 0 0 0, clip, width=0.27\linewidth]{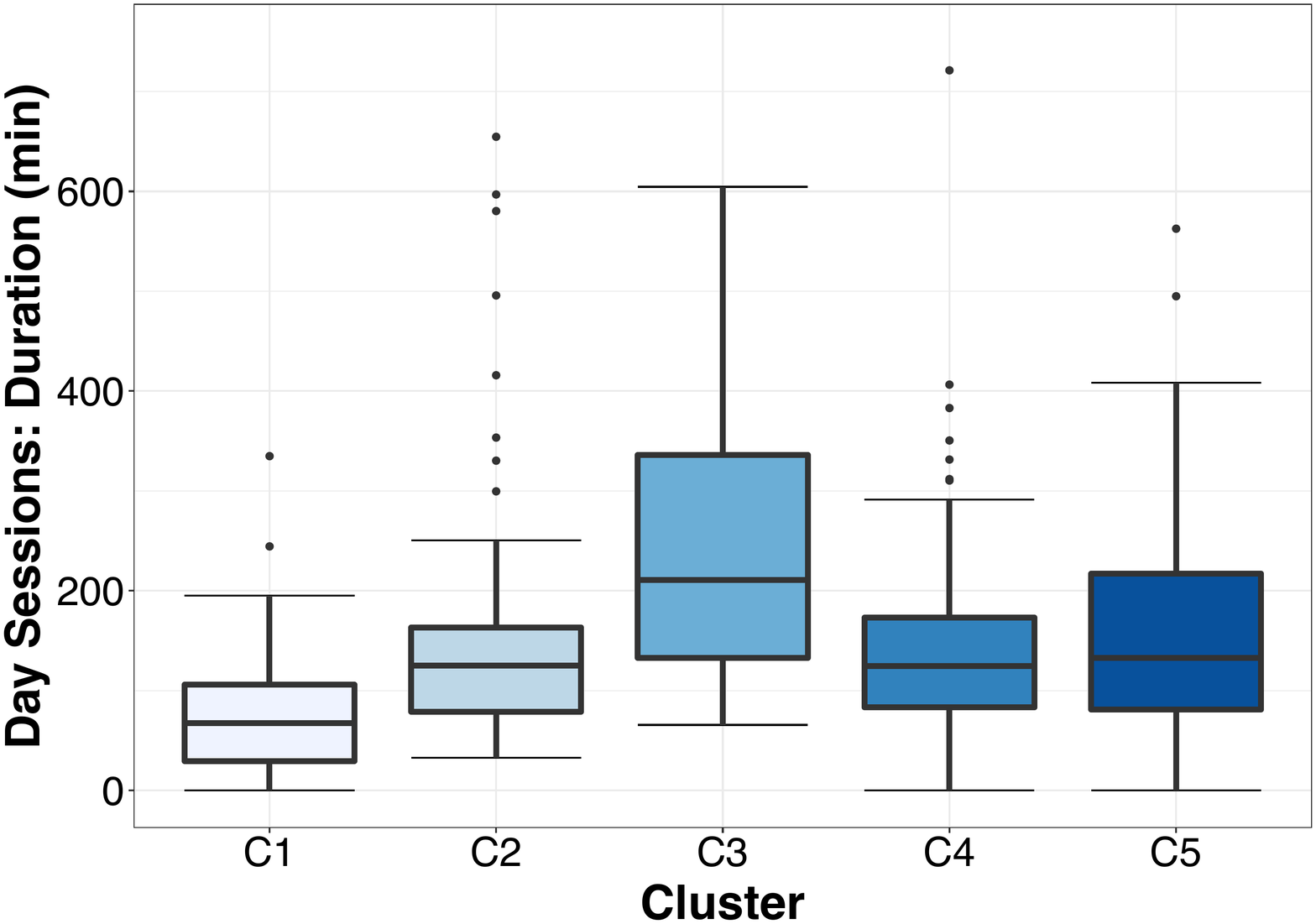}\label{feature-2}
\hfil
\includegraphics[trim=0 0 0 0, clip, width=0.27\linewidth]{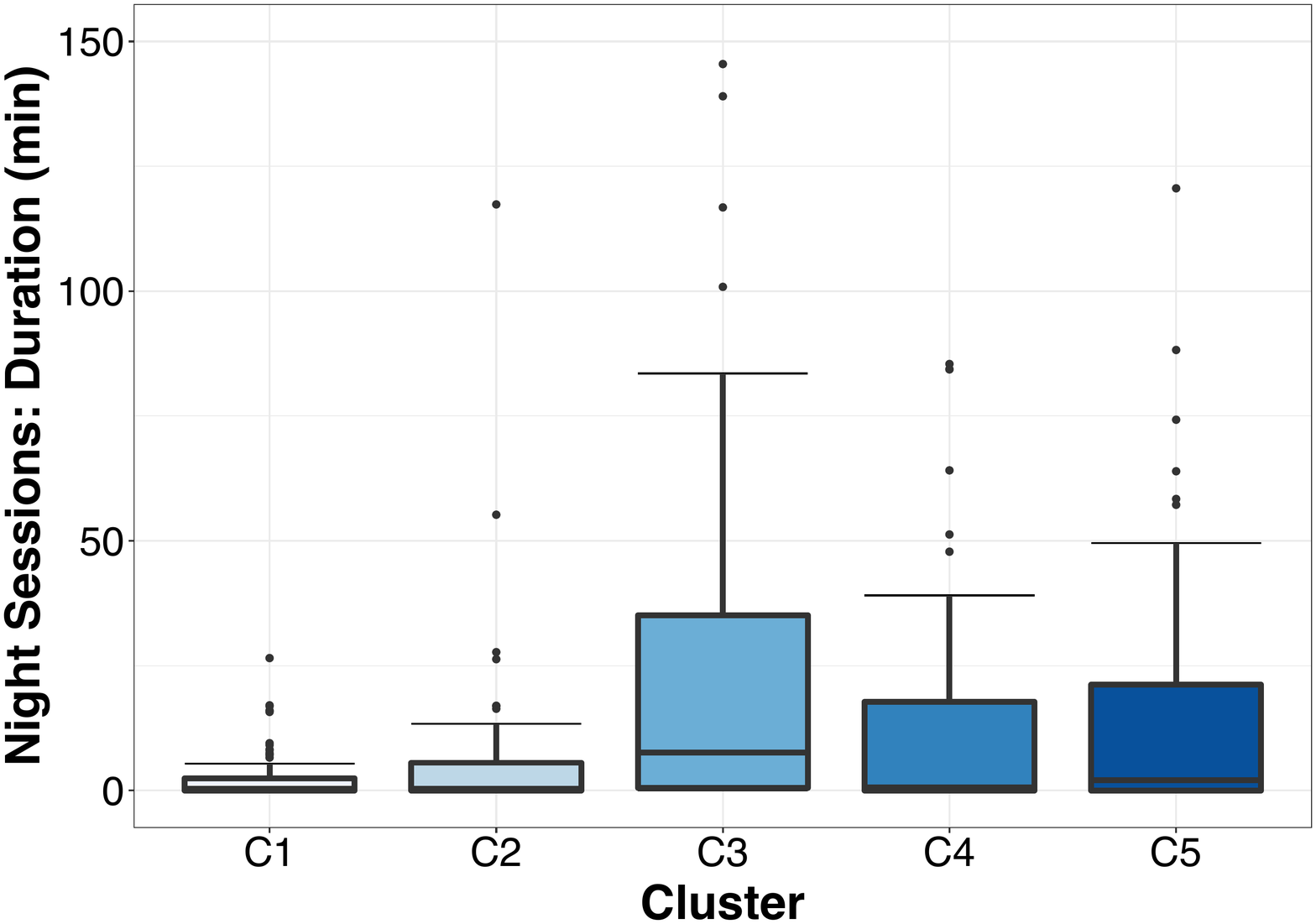}\label{feature-3}
}
\vspace{0.2cm}\\
\hbox{
\includegraphics[trim=0 0 0 0, clip, width=0.27\linewidth]{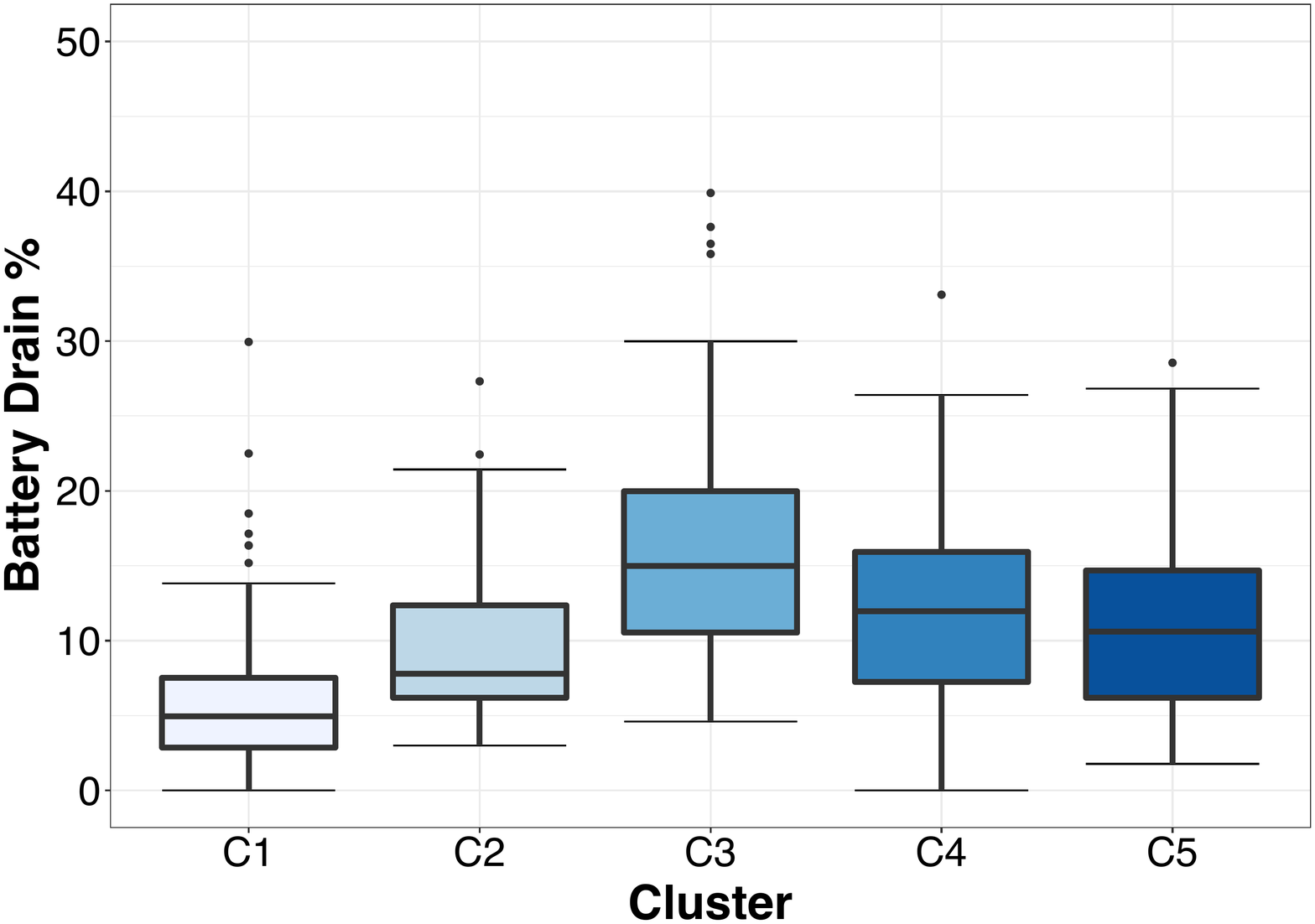}\label{feature-4}
\hfil
\includegraphics[trim=0 0 0 0, clip, width=0.27\linewidth]{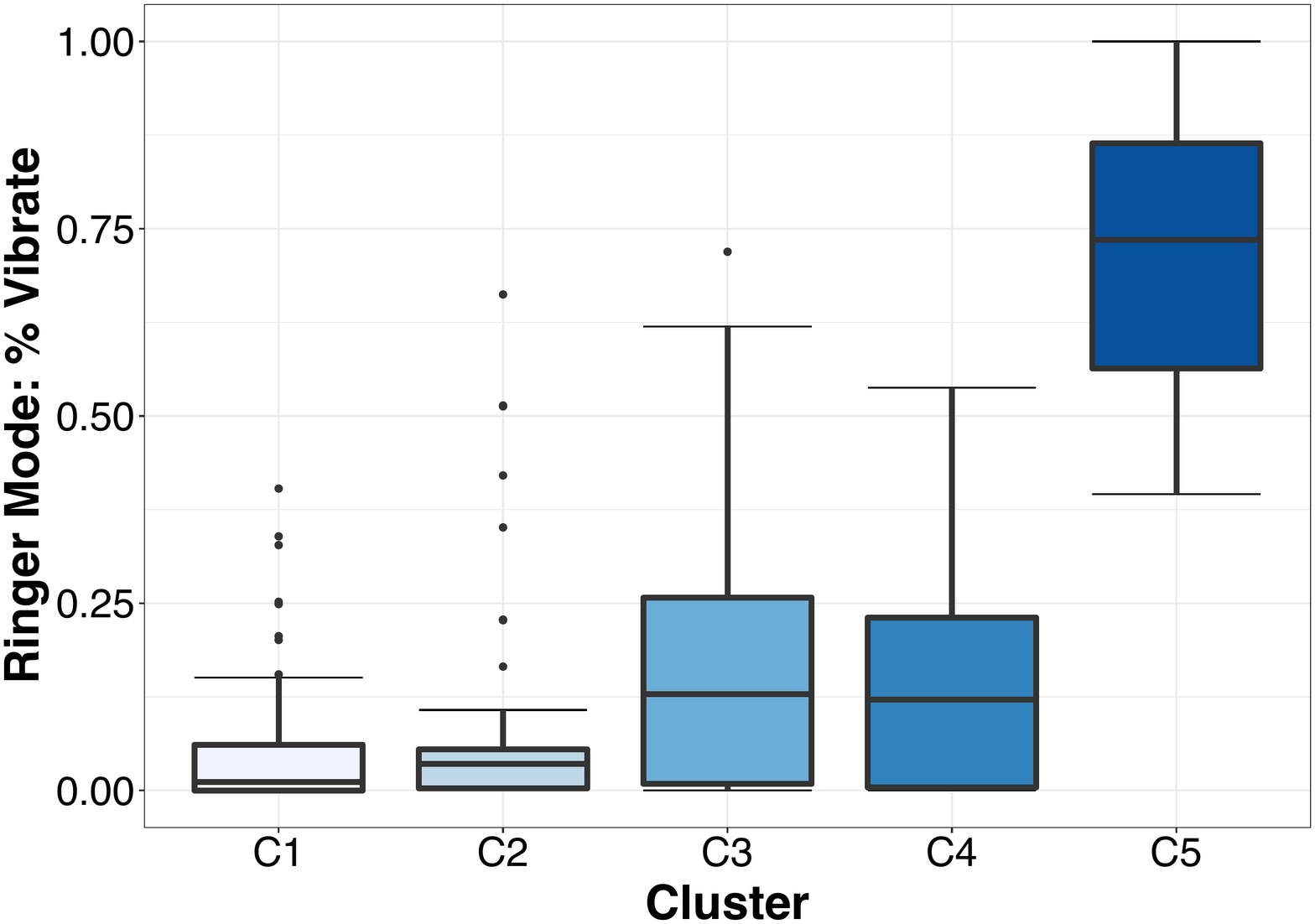}\label{feature-5}
\hfil
\includegraphics[trim=0 0 0 0, clip, width=0.27\linewidth]{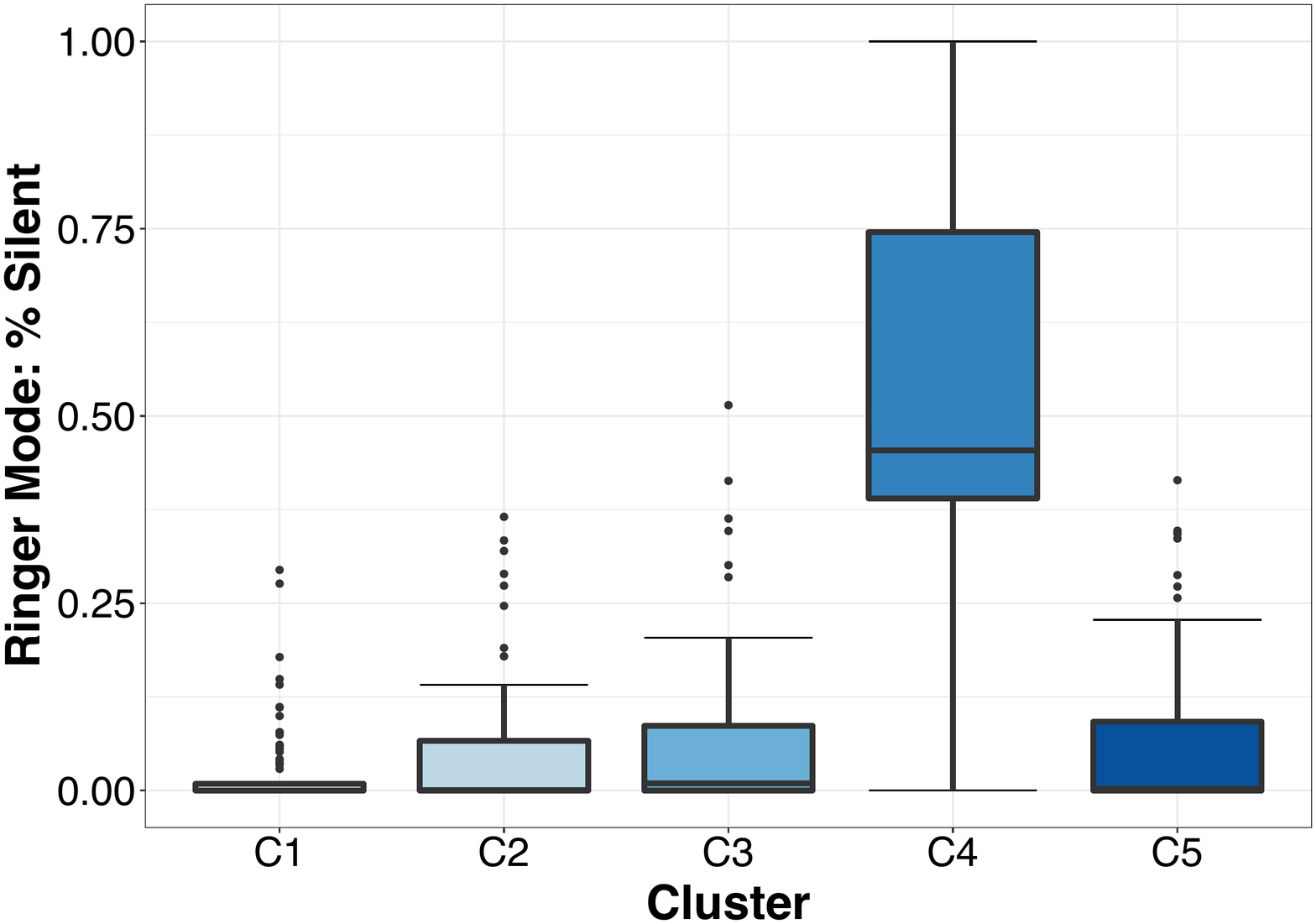}\label{feature-6}
}
\end{tabular}
\caption{Examples of score distributions of highly discriminatory features across predicted clusters} 
\label{fig:features}
\vspace{-0.3cm}
\end{figure*}

For our repeated measures data (\emph{e.g.}, ESM), the next step for progressing from a regression model to a growth model (via model comparisons) was to establish a simple model without any random effects to serve as a baseline. This step involved modeling the fixed relationship between time and the corresponding dependent variable. We began by fitting a linear relationship and progressively added more complicated relationships such as quadratic, cubic, etc. The results indicated whether time is significantly related to the dependent variable.
 
A potential limitation of the previous model is that it assumes that the relationship between time and the dependent variable is constant for all individuals. An alternative model is one that allows slopes to randomly vary. Therefore, a random slope model was tested by adding the linear effect of time as a random effect. The results of this analysis indicated whether a model that allows the slope between time and the dependent variable to randomly vary fits the data better than a model that fixes the slope to a constant value for all individuals.

The fourth step in developing the level-1 model involved assessing the error structure of the model since significance tests may be dramatically affected if error structures are not properly specified. Therefore, we determined whether a model fit improves by incorporating (a) an autoregressive structure with serial correlations and (b) heterogeneity in the error structures.

\begin{table}[t!]
	\scriptsize
	\caption{Summary of the models for the ESM:Tense-Calm scale}
	\label{tbl:esm-tense-calm}
	\centering
	\begin{tabular}{@{} L{31mm}dddd @{}}
		\toprule
			& \multicolumn{1}{C{10mm}}{\textbf{Model 2~~~}} & \multicolumn{1}{C{10mm}}{\textbf{Model 3~~~}} & \multicolumn{1}{C{10mm}}{\textbf{Model 4~~~}} & \multicolumn{1}{C{10mm}}{\textbf{Model 5~~~}}	\\
		\midrule
		\midrule
		\textit{Fixed factors}	                        &	                        &	                        &	                        &	\\
		\hspace{2mm}Intercept                       	&	$-.59$\rlap{$^{***}$}	&	$-.59$\rlap{$^{***}$}	&	$-.57$\rlap{$^{***}$}	&	$-.57$\rlap{$^{***}$}	\\
		\hspace{2mm}Cluster:2                       	&	$.07$	                &	$.07$	                &	$.09$               	&	$.09$	\\
		\hspace{2mm}Cluster:3                       	&	$.00$	                &	$.00$  	                &	$-.00$              	&	$-.00$	\\
		\hspace{2mm}Cluster:4                          	&	$.11$\rlap{$^{*}$}     	&	$.12$\rlap{$^{*}$}  	&	$.12$\rlap{$^{*}$} 	    &	$.13$\rlap{$^{*}$}	\\
		\hspace{2mm}Cluster:5                       	&	$.11$\rlap{$^{*}$}      &	$.12$\rlap{$^{*}$}      &	$.14$\rlap{$^{*}$}     	&	$.14$\rlap{$^{*}$}	\\
		\midrule
		\hspace{2mm}isNight:Yes	                        &	-                   	&	$-.01$	                &	-	                    &	$-.01$	\\
		\hspace{2mm}Cluster:2 $\times$ isNight:Yes  	&	-                       &	$.01$	                &	-                   	&	$.02$	\\
		\hspace{2mm}Cluster:3 $\times$ isNight:Yes  	&	-                       &	$-.02$	                &	-                   	&	$-.02$	\\
		\hspace{2mm}Cluster:4 $\times$ isNight:Yes	    &	-                       &	$-.06$	                &	-	                    &	$-.06$	\\
		\hspace{2mm}Cluster:5 $\times$ isNight:Yes	    &	-                       &	$-.01$	                &	-               	    &	$-.01$	\\
		\midrule
		\hspace{2mm}WorkingDay:No	                    &	-  	                    &	-	                    &	$-.05$\rlap{$^{***}$}	&	$-.05$\rlap{$^{***}$}	\\
		\hspace{2mm}Cluster:2 $\times$ WorkingDay:No	&	-                       &	-	                    &	$-.07$\rlap{$^{**}$}	&	$-.07$\rlap{$^{**}$}	\\
		\hspace{2mm}Cluster:3 $\times$ WorkingDay:No	&	-                       &	-	                    &	$.03$	                &	$.03$	\\
		\hspace{2mm}Cluster:4 $\times$ WorkingDay:No	&	-                       &	-	                    &	$-.04$	                &	$-.04$	\\
		\hspace{2mm}Cluster:5 $\times$ WorkingDay:No	&	-                       &	-	                    &	$-.07$\rlap{$^{**}$}	&	$-.07$\rlap{$^{**}$}	\\
		\midrule
		\textit{Random factors}                     	&	                        &	                        &	                        &	\\
		\hspace{2mm}Intercept	                        &	$.35$   	            &	$.35$	                &	$.35$       	        &	$.35$	\\
		\hspace{2mm}Residuals   	                    &	$.54$	                &	$.54$	                &	$.54$	                &	$.54$	\\
		\midrule
        \multicolumn{5}{l}{No. Observations: $31396$; No. Groups: $340$}   \\
		\bottomrule
		\multicolumn{5}{l}{$^{*}p~<~.05;~^{**}p~<~.01;~^{***}p~<~.001$} \\
	\end{tabular}
	\vspace{-0.3cm}
\end{table}

\begin{table}[t!]
	\scriptsize
	\caption{Summary of the models for the ESM:Tired-Awake scale}
	\label{tbl:esm-tired-awake}
	\centering
	\begin{tabular}{@{} L{31mm}dddd @{}}
		\toprule
			& \multicolumn{1}{C{10mm}}{\textbf{Model 2~~~}} & \multicolumn{1}{C{10mm}}{\textbf{Model 3~~~}} & \multicolumn{1}{C{10mm}}{\textbf{Model 4~~~}} & \multicolumn{1}{C{10mm}}{\textbf{Model 5~~~}}	\\
		\midrule
		\midrule
		\textit{Fixed factors}	                        &	                        &	                        &	                        &	\\
		\hspace{2mm}Intercept                       	&	$.28$\rlap{$^{***}$}	&	$.31$\rlap{$^{***}$}	&	$.26$\rlap{$^{***}$}	&	$.28$\rlap{$^{***}$}	\\
		\hspace{2mm}Cluster:2                       	&	$-.11$	                &	$-.12$	                &	$-.11$               	&	$-.12$	\\
		\hspace{2mm}Cluster:3                       	&	$-.05$	                &	$-.03$  	            &	$-.03$              	&	$-.01$	\\
		\hspace{2mm}Cluster:4                          	&	$-.12$              	&	$-.10$    	            &	$-.11$	                &	$-.10$	\\
		\hspace{2mm}Cluster:5                       	&	$-.13$\rlap{$^{*}$}     &	$-.12$                  &	$-.13$\rlap{$^{*}$}     &	$-.12$	\\
		\midrule
		\hspace{2mm}isNight:Yes	                        &	-                   	&	$-.58$\rlap{$^{***}$}   &	-	                    &	$-.58$\rlap{$^{***}$}	\\
		\hspace{2mm}Cluster:2 $\times$ isNight:Yes  	&	-                       &	$.10$	                &	-                   	&	$.09$	\\
		\hspace{2mm}Cluster:3 $\times$ isNight:Yes  	&	-                       &	$.03$	                &	-                   	&	$.03$	\\
		\hspace{2mm}Cluster:4 $\times$ isNight:Yes	    &	-                       &	$.08$	                &	-	                    &	$.08$	\\
		\hspace{2mm}Cluster:5 $\times$ isNight:Yes	    &	-                       &	$.12$\rlap{$^{**}$}     &	-               	    &	$.12$\rlap{$^{*}$}	\\
		\midrule
		\hspace{2mm}WorkingDay:No	                    &	-  	                    &	-	                    &	$.09$\rlap{$^{***}$}	&	$.09$\rlap{$^{***}$}	\\
		\hspace{2mm}Cluster:2 $\times$ WorkingDay:No	&	-                       &	-	                    &	$-.00$	                &	$.00$\rlap{$^{*}$}	\\
		\hspace{2mm}Cluster:3 $\times$ WorkingDay:No	&	-                       &	-	                    &	$-.06$\rlap{$^{**}$}    &	$-.05$	\\
		\hspace{2mm}Cluster:4 $\times$ WorkingDay:No	&	-                       &	-	                    &	$-.01$	                &	$-.01$	\\
		\hspace{2mm}Cluster:5 $\times$ WorkingDay:No	&	-                       &	-	                    &	$-.01$	                &	$.01$	\\
		\midrule
		\textit{Random factors}                     	&	                        &	                        &	                        &	\\
		\hspace{2mm}Intercept	                        &	$.39$   	            &	$.39$	                &	$.39$       	        &	$.39$	\\
		\hspace{2mm}Residuals   	                    &	$.68$	                &	$.67$	                &	$.68$	                &	$.67$	\\
		\midrule
        \multicolumn{5}{l}{No. Observations: $31373$; No. Groups: $340$}   \\

		\bottomrule
		\multicolumn{5}{l}{$^{*}p~<~.05;~^{**}p~<~.01;~^{***}p~<~.001$} \\
	\end{tabular}
	\vspace{-0.3cm}
\end{table}

For building the level-2 model we considered for each of the growth parameters that contain meaningful variability (\emph{e.g.}, intercept, linear) the appropriate individual difference predictors, as well as their interactions. More specifically, we fitted a series of models for each mood and psychometric scale, while other factors were added as predictors, to assess if the magnitude of each scale is related to the cluster assignment when controlling for the effects of those predictors. Besides the cluster assignment independent variable (Model 2), the models accounted for the time window (day vs. night) of the observations (Model 3), whether its a working day (Model 4), as well as all the aforementioned predictors and their interactions altogether (Model 5).

\section{A Proposed Taxonomy of Mobile Phone Users}

Table~\ref{tbl:descriptive_statistics} shows the means and standard deviations for the features we used to derive the clusters of mobile phone users. As a reference, we also provide the same descriptive statistics for all users in our study.

\begin{table}[t!]
	\scriptsize
	\caption{Summary of the models for the ESM:Valence scale}
	\label{tbl:esm-valence}
	\centering
	\begin{tabular}{@{} L{31mm}dddd @{}}
		\toprule
			& \multicolumn{1}{C{10mm}}{\textbf{Model 2~~~}} & \multicolumn{1}{C{10mm}}{\textbf{Model 3~~~}} & \multicolumn{1}{C{10mm}}{\textbf{Model 4~~~}} & \multicolumn{1}{C{10mm}}{\textbf{Model 5~~~}}	\\
		\midrule
		\midrule
		\textit{Fixed factors}	                        &	                        &	                        &	                        &	\\
		\hspace{2mm}Intercept                       	&	$.74$\rlap{$^{***}$}	&	$.74$\rlap{$^{***}$}    &	$.74$\rlap{$^{***}$}    &	$.74$\rlap{$^{***}$}	\\
		\hspace{2mm}Cluster:2                       	&	$-.02$	                &	$-.02$	                &	$-.02$               	&	$-.03$	\\
		\hspace{2mm}Cluster:3                       	&	$-.05$	                &	$-.05$  	            &	$-.04$              	&	$-.04$	\\
		\hspace{2mm}Cluster:4                          	&	$-.11$\rlap{$^{*}$}     &	$-.10$\rlap{$^{*}$}     &	$-.11$\rlap{$^{*}$}     &	$-.11$\rlap{$^{*}$}	\\
		\hspace{2mm}Cluster:5                       	&	$-.11$\rlap{$^{*}$}     &	$-.11$\rlap{$^{*}$}     &	$-.13$\rlap{$^{*}$}     &	$-.12$\rlap{$^{*}$}	\\
		\midrule
		\hspace{2mm}isNight:Yes	                        &	-                   	&	$-.04$                  &	-	                    &	$-.04$\rlap{$^{*}$}	\\
		\hspace{2mm}Cluster:2 $\times$ isNight:Yes  	&	-                       &	$.00$	                &	-                   	&	$.00$	\\
		\hspace{2mm}Cluster:3 $\times$ isNight:Yes  	&	-                       &	$.01$	                &	-                   	&	$.01$	\\
		\hspace{2mm}Cluster:4 $\times$ isNight:Yes	    &	-                       &	$.01$	                &	-	                    &	$.01$	\\
		\hspace{2mm}Cluster:5 $\times$ isNight:Yes	    &	-                       &	$.00$                   &	-               	    &	$.00$	\\
		\midrule
		\hspace{2mm}WorkingDay:No	                    &	-  	                    &	-	                    &	$.00$               	&	$.00$	\\
		\hspace{2mm}Cluster:2 $\times$ WorkingDay:No	&	-                       &	-	                    &	$.02$	                &	$.02$   	\\
		\hspace{2mm}Cluster:3 $\times$ WorkingDay:No	&	-                       &	-	                    &	$-.01$                  &	$-.01$	\\
		\hspace{2mm}Cluster:4 $\times$ WorkingDay:No	&	-                       &	-	                    &	$.02$	                &	$.02$	\\
		\hspace{2mm}Cluster:5 $\times$ WorkingDay:No	&	-                       &	-	                    &	$.04$\rlap{$^{*}$}      &	$.04$\rlap{$^{*}$}	\\
		\midrule
		\textit{Random factors}                     	&	                        &	                        &	                        &	\\
		\hspace{2mm}Intercept	                        &	$.32$   	            &	$.32$	                &	$.32$       	        &	$.32$	\\
		\hspace{2mm}Residuals   	                    &	$.44$	                &	$.44$	                &	$.44$	                &	$.44$	\\
		\midrule
        \multicolumn{5}{l}{No. Observations: $31789$; No. Groups: $340$}   \\
		\bottomrule
		\multicolumn{5}{l}{$^{*}p~<~.05;~^{**}p~<~.01;~^{***}p~<~.001$} \\
	\end{tabular}
	\vspace{-0.3cm}
\end{table}

\begin{table}[t!]
	\scriptsize
	\caption{Summary of the models for the ESM:Boredom scale}
	\label{tbl:esm-boredom}
	\centering
	\begin{tabular}{@{} L{31mm}dddd @{}}
		\toprule
			& \multicolumn{1}{C{10mm}}{\textbf{Model 2~~~}} & \multicolumn{1}{C{10mm}}{\textbf{Model 3~~~}} & \multicolumn{1}{C{10mm}}{\textbf{Model 4~~~}} & \multicolumn{1}{C{10mm}}{\textbf{Model 5~~~}}	\\
		\midrule
		\midrule
		\textit{Fixed factors}	                        &	                        &	                        &	                        &	\\
		\hspace{2mm}Intercept                       	&	$.09$\rlap{$^{***}$}	&	$.09$\rlap{$^{***}$}    &	$.09$\rlap{$^{***}$}    &	$.08$\rlap{$^{***}$}	\\
		\hspace{2mm}Cluster:2                       	&	$-.00$	                &	$-.00$	                &	$-.00$               	&	$-.00$	\\
		\hspace{2mm}Cluster:3                       	&	$.02$	                &	$.03$  	                &	$.03$                   &	$.03$	\\
		\hspace{2mm}Cluster:4                          	&	$.05$\rlap{$^{*}$}      &	$.05$\rlap{$^{*}$}      &	$.04$	                &	$.04$	\\
		\hspace{2mm}Cluster:5                       	&	$.06$\rlap{$^{**}$}     &	$.06$\rlap{$^{**}$}     &	$.06$\rlap{$^{**}$}     &	$.06$\rlap{$^{**}$}	\\
		\midrule
		\hspace{2mm}isNight:Yes	                        &	-                   	&	$.03$\rlap{$^{**}$}     &	-	                    &	$.03$\rlap{$^{**}$} 	\\
		\hspace{2mm}Cluster:2 $\times$ isNight:Yes  	&	-                       &	$-.02$	                &	-                   	&	$-.03$	\\
		\hspace{2mm}Cluster:3 $\times$ isNight:Yes  	&	-                       &	$-.04$\rlap{$^{**}$}    &	-                   	&	$-.04$\rlap{$^{**}$}	\\
		\hspace{2mm}Cluster:4 $\times$ isNight:Yes	    &	-                       &	$-.04$\rlap{$^{**}$}    &	-	                    &	$-.04$\rlap{$^{**}$}	\\
		\hspace{2mm}Cluster:5 $\times$ isNight:Yes	    &	-                       &	$.00$                   &	-               	    &	$-.00$	\\
		\midrule
		\hspace{2mm}WorkingDay:No	                    &	-  	                    &	-	                    &	$.01$               	&	$.01$	\\
		\hspace{2mm}Cluster:2 $\times$ WorkingDay:No	&	-                       &	-	                    &	$.03$	                &	$.03$\rlap{$^{**}$}   	\\
		\hspace{2mm}Cluster:3 $\times$ WorkingDay:No	&	-                       &	-	                    &	$-.00$                  &	$-.00$	\\
		\hspace{2mm}Cluster:4 $\times$ WorkingDay:No	&	-                       &	-	                    &	$.03$	                &	$.03$\rlap{$^{**}$}	\\
		\hspace{2mm}Cluster:5 $\times$ WorkingDay:No	&	-                       &	-	                    &	$.00$	                &	$.00$	\\
		\midrule
		\textit{Random factors}                     	&	                        &	                        &	                        &	\\
		\hspace{2mm}Intercept	                        &	$.14$   	            &	$.14$	                &	$.14$       	        &	$.14$	\\
		\hspace{2mm}Residuals   	                    &	$.26$	                &	$.26$	                &	$.26$	                &	$.26$	\\
		\midrule
        \multicolumn{5}{l}{No. Observations: $32307$; No. Groups: $340$}   \\
		\bottomrule
		\multicolumn{5}{l}{$^{*}p~<~.05;~^{**}p~<~.01;~^{***}p~<~.001$} \\
	\end{tabular}
	\vspace{-0.2cm}
\end{table}

In this section, we examine the characteristics of each cluster and draw insights on how different user profiles interact with mobile phone technology based on the clusters' structural description (selected examples of cluster variations are shown in Figure~\ref{fig:features}). Following the findings of our multilevel modelling (presented in Tables~\ref{tbl:esm-tense-calm}-\ref{tbl:Big5-Neuroticism}), we also highlight potential implications on the emotional well-being of mobile phone users. Note that we report only statistically significant findings at $p < .05$, $p < .01$ and $p < .001$, for which we have corrected the level of significance using the Holm's~\cite{Holm:1979} method. For data that meet the parametric assumptions we applied the one-way independent ANOVA test, whereas for all other cases we used the Kruskal-Wallis test. When a difference was found to be significant, we followed that finding by applying the independent t-test or the Wilcoxon signed-rank test, respectively.
Considering the significant differences across the clusters, we now conclude to a set of five user profiles.

\subsection{Cluster 1 -- Limited Use (Baseline)}
99 participants fall into this cluster, with 54.55\% of its members being female and 45.45\% male. Their age ranges from 18 to 66 ($M = 42.4$, $SD = 10.2$).

In comparison to the other clusters, its members score low in almost all usage categories. They tend to keep their ringer mode in the normal setting, a behaviour that might be related to not being bothered too much by incoming calls and notifications.

Since this cluster shows the lowest amount of phone use, we use it as a baseline for the other clusters to compare the impact that different forms of intensified use have on emotional well-being.

\begin{table}[t!]
	\scriptsize
	\caption{Summary of the models for the PHQ-8 scale}
	\label{tbl:PHQ8}
	\centering
	\begin{tabular}{@{} L{16mm}ddddd @{}}
		\toprule
			& \multicolumn{1}{C{10mm}}{\textbf{C1~~~~~}} & \multicolumn{1}{C{10mm}}{\textbf{C2~~~~~}} & \multicolumn{1}{C{10mm}}{\textbf{C3~~~~~}} & \multicolumn{1}{C{10mm}}{\textbf{C4~~~~~}} & \multicolumn{1}{C{10mm}}{\textbf{C5~~~~~}}	\\
		\midrule
		\midrule
		\textit{Fixed factors}	                        &	                        &	                        &	                        &                           &   \\
		\hspace{2mm}Intercept                       	&	$5.40$\rlap{$^{***}$}	&	$8.50$\rlap{$^{***}$}   &	$6.85$\rlap{$^{***}$}   &	$10.50$\rlap{$^{***}$}  &   $9.00$\rlap{$^{***}$}   \\
		\hspace{2mm}Cluster:1                       	&	-	                    &	$-3.09$	                &	$-1.44$               	&	$-5.09$\rlap{$^{**}$}   &   $-3.59$\rlap{$^{*}$}	\\
		\hspace{2mm}Cluster:2                       	&	$3.09$	                &	-	                    &	$1.64$               	&	$-2.00$                 &   $-.50$	\\
		\hspace{2mm}Cluster:3                       	&	$1.44$	                &	$-1.64$  	            &	-              	        &	$-3.64$\rlap{$^{*}$}    &   $-2.14$	\\
		\hspace{2mm}Cluster:4                          	&	$5.09$\rlap{$^{**}$}    &	$2.00$                  &	$3.64$\rlap{$^{**}$}    &	-                       &   $1.50$	\\
		\hspace{2mm}Cluster:5                       	&	$3.59$\rlap{$^{*}$}     &	$.50$                   &	$2.14$                  &	$-1.50$                 &   -	\\
		\midrule
		\textit{Random factors}                     	&	                        &	                        &	                        &                           &	\\
		\hspace{2mm}Intercept	                        &	$4.11$   	            &	$3.96$	                &	$4.11$       	        &	$4.11$                  &	$4.11$	\\
		\midrule
        \multicolumn{5}{l}{No. Observations: $65$; No. Groups: $63$}   \\
		\bottomrule
		\multicolumn{6}{l}{$^{*}p~<~.05;~^{**}p~<~.01;~^{***}p~<~.001$} \\
	\end{tabular}
	\vspace{-0.4cm}
\end{table}

\begin{figure}[t!]
\vspace{0.1cm}
\centering
\includegraphics[trim=0 0 0 0, clip, width=0.5\linewidth]{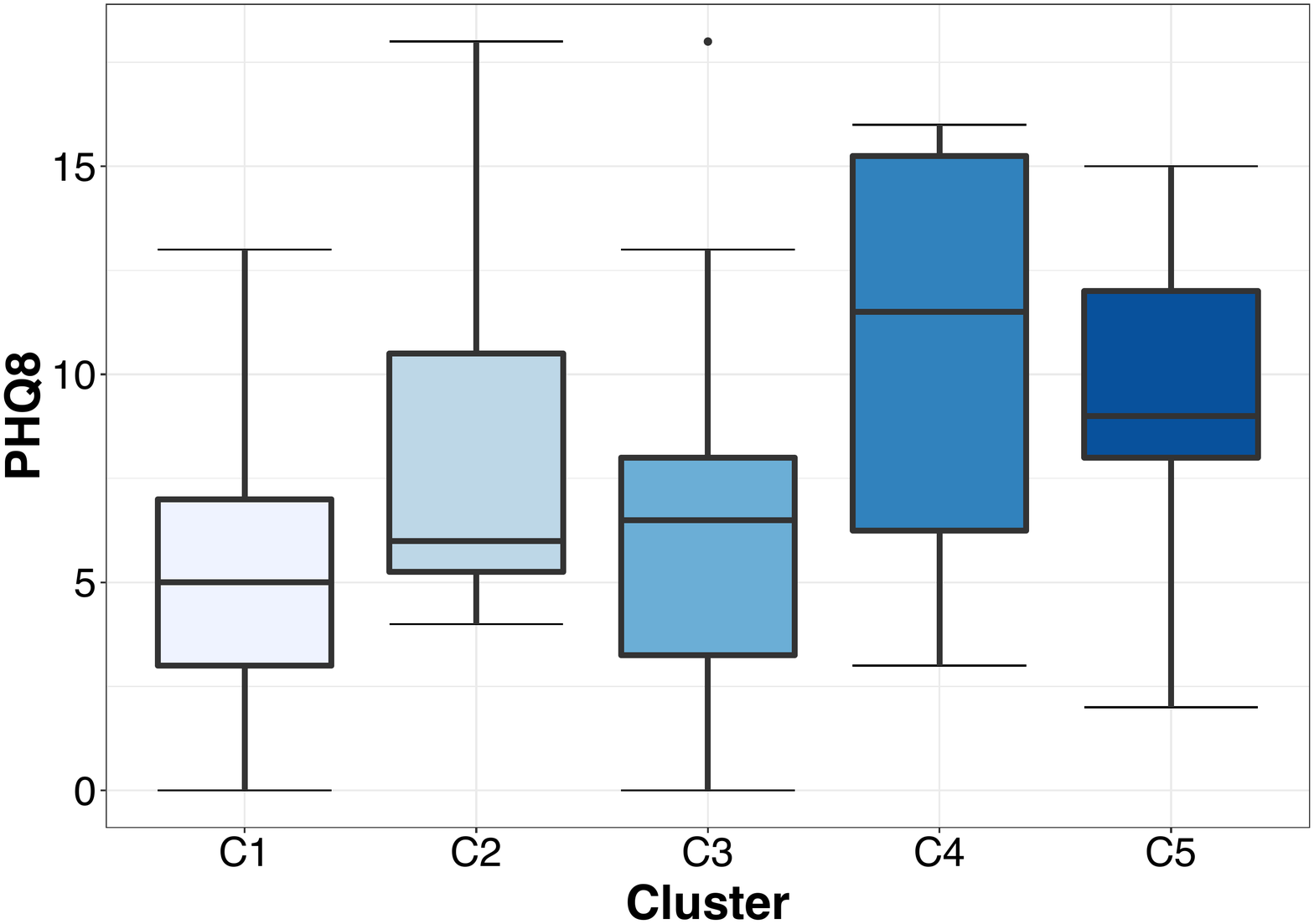}
\caption{PHQ-8 scores per cluster}
\label{fig:PHQ8}
\vspace{-0.4cm}
\end{figure}

\begin{table}[t]
	\scriptsize
	\caption{Summary of the models for the Big5: Neuroticism scale}
	\label{tbl:Big5-Neuroticism}
	\centering
	\begin{tabular}{@{} L{16mm}ddddd @{}}
		\toprule
			& \multicolumn{1}{C{10mm}}{\textbf{C1~~~~~}} & \multicolumn{1}{C{10mm}}{\textbf{C2~~~~~}} & \multicolumn{1}{C{10mm}}{\textbf{C3~~~~~}} & \multicolumn{1}{C{10mm}}{\textbf{C4~~~~~}} & \multicolumn{1}{C{10mm}}{\textbf{C5~~~~~}}	\\
		\midrule
		\midrule
		\textit{Fixed factors}	                        &	                        &	                        &	                        &                           &   \\
		\hspace{2mm}Intercept                       	&	$10.95$\rlap{$^{***}$}	&	$11.22$\rlap{$^{***}$}  &	$12.14$\rlap{$^{***}$}  &	$9.88$\rlap{$^{***}$}   &   $13.58$\rlap{$^{***}$}   \\
		\hspace{2mm}Cluster:1                       	&	-   	                &	$-.27$	                &	$-1.19$               	&	$1.06$                  &   $-2.63$	\\
		\hspace{2mm}Cluster:2                       	&	$.27$	                &	-   	                &	$-.92$               	&	$1.33$                  &   $-2.35$	\\
		\hspace{2mm}Cluster:3                       	&	$1.19$	                &	$.92$  	                &	-                     	&	$2.25$\rlap{$^{*}$}     &   $-1.43$	\\
		\hspace{2mm}Cluster:4                          	&	$-1.06$                 &	$-1.33$                 &	$-2.25$\rlap{$^{*}$}    &	-                       &   $-3.70$\rlap{$^{**}$}	\\
		\hspace{2mm}Cluster:5                       	&	$2.63$\rlap{$^{**}$}    &	$2.35$                  &	$1.43$                  &	$3.70$\rlap{$^{**}$}    &   -   	\\
		\midrule
		\textit{Random factors}                     	&	                        &	                        &	                        &                           &	\\
		\hspace{2mm}Intercept	                        &	$4.22$   	            &	$4.22$	                &	$4.22$       	        &	$4.22$                  &	$4.22$	\\
		\midrule
        \multicolumn{5}{l}{No. Observations: $166$; No. Groups: $164$}   \\
		\bottomrule
		\multicolumn{6}{l}{$^{*}p~<~.05;~^{**}p~<~.01;~^{***}p~<~.001$} \\
	\end{tabular}
\end{table}

\vspace{-0.1cm}

\subsection{Cluster 2 -- Business Use}
45 participants fall into the Business Use cluster, where 44.44\% of its members are female and 55.56\% are male. Their age ranges from 26 to 62 ($M = 43.2$, $SD = 8.7$).

Members of this cluster stand out by their significantly more frequent use of phone calls -- incoming and outgoing. While phone calls are comparably frequent, members of this cluster have comparably fewer nightly use sessions and app launches. The ringer mode is typically set to normal, indicating that hearing the phone is important to them. Even though such phone use patterns could be seen in several different user categories, we believe that these patterns are indicative of business use, hence we used such label.

In terms of well-being, members of this cluster did not stand out much. Only during the weekend, they were found to report significantly lower tense arousal. When accounting for night-time and day of the week, we found that during the weekend, members of the cluster reported higher levels of boredom than the baseline cluster.

\subsection{Cluster 3 -- Power Use}
67 participants fall into the Power Use cluster, with 56.72\% of the members being female and 43.28\% male. Their age ranges from 19 to 56 ($M = 37.4$, $SD = 9.2$).

Members of this cluster stand out by their increased session duration \& number of nightly sessions, battery use, and mobile data use. During the day, they launched a significantly higher number of email-, game-, and to a lesser extent social media apps. Messaging apps, in contrast, are used less often during the day compared to other clusters. During the night, members of this cluster show the highest use of email apps and a somewhat increase use of messaging apps. While the ringer mode is typically set to normal mode, members of this cluster had the highest variance in ringer mode setting.

Despite the high level of mobile phone use, members of the Power Use cluster do not stand out negatively in any of the well-being related factors. Compared to the baseline, they are more awake during the weekend and report lower levels of boredom during the night. Compared to Cluster 4 mentioned below, they scored lower in terms of depression (PHQ-8) and neuroticism (Big5).

\subsection{Cluster 4 -- Personality-Induced Problematic Phone Use}
62 participants fall into this cluster, with 58.06\% of its members being female and 41.94\% male. Their age ranges from 18 to 62 ($M = 33.1$, $SD = 11.9$).

Members of this cluster stand out by an increased number and length of sessions during night time, and an increased use of email- and messaging apps during the night. The ringer is typically set to silent mode.
Members of this cluster scored worst in terms of well-being. Compared to the baseline cluster, they reported significantly higher levels of tense-arousal, boredom, and lower valence. In contrast, when accounting for night-time and non-working days, the significant differences regarding tense-arousal and valence disappeared. Bored, however, was significantly lower during night-time. These findings indicate a tendency towards experiencing more stress and boredom during working hours.
Members of this cluster further tended to be more neurotic / less emotionally stable than members of other clusters.
Finally, members of this cluster scored significantly higher on the PHQ-8 questionnaire than Limited and Power Users, indicating a tendency towards experiencing depression-related symptoms.

\subsection{Cluster 5 -- Externally-Induced Problematic Phone Use}
67 participants fall into this cluster, with 49.25\% of its members being female and 50.75\% male. Their age ranges from 18 to 58 $(M = 32.4$, $SD = 10.0$).

Like the previous cluster, members of this cluster tend to have more and longer phone use sessions during night-time. The main difference to Cluster~4 is that during night-time, only the use of messaging apps is comparably higher. In contrast, the use of email apps is lower. Finally, the ringer mode of these users is typically set to vibrate.

There were significant effects of membership of this cluster on the emotional self-reports. Compared to the baseline cluster, its members reported significantly higher tense arousal, lower energetic arousal, lower valence, and higher levels of boredom.
During night-time, however, energetic arousal was significantly lower -- an indication of being more tired -- and the other effects subsided. During the weekend, tense arousal was significantly lower, valence was significantly higher, and significant effects on energetic arousal and boredom disappeared.

Also, members of this cluster scored significantly higher in terms of emotional stability compared to the baseline cluster as well as Cluster~4. Further, PHQ-8 depression scores (Figure~\ref{fig:PHQ8}) were significantly higher than those of the baseline cluster, however, the effect was not as pronounced as in Cluster~4.

While members of this cluster tend to be stressed during working time, they seem to better compensate during non-working hours: they are being tired during the night and happy during the weekend. This finding is corroborated by higher emotional stability. We interpret the main difference that members of this cluster have a stressful daytime, but are more affected by external factors rather than internal factors, and therefore better cope with the stressful weekdays.

\section{Discussion}

We identified five clusters of phone use: Limited Use, Business Use, Power Use, Personality-Induced Problematic Phone Use, and Externally-Induced Problematic Phone Use. We used the Limited Use cluster as baseline to compare the other clusters against. Business Users tend to use the mobile phone as phone primarily. Power Users use extensively the smartphone capabilities. Personality-Induced Problematic Phone Use and Externally-Induced Problematic Phone Use both stand out by their night sessions and not having the ringer in normal mode.

\subsection{Heavy phone use does not predict negative well-being}
Surprisingly, we found that extensive use of the phone alone did not predict negative emotional well-being. Neither Business Users nor Power Users, who scored highest in phone calling and session duration, respectively, scored negatively in terms of affect or personality. Only members of the Clusters 4 and 5 -- who were using the phone overall less than Power Users -- scored significantly higher on the PHQ-8 depression scale. Thus, rather than using the phone a lot, it's comparably frequent and long night-time use that predicts negative effects on emotional states. Members of the Clusters 4 and 5 -- those with more and longer night-time sessions -- felt tenser, worse, and more bored, in particular during working days.

This finding contrasts with previous work. For example, Thomee \emph{et al.}~\cite{Thomee:2011} found that high mobile phone use was associated with sleep disturbances and symptoms of depression in Sweden. Yen \emph{et al.}~\cite{Yen:2009} also reported an association of high mobile phone use, outgoing phone calls and text messages with depression.
However, our results indicate that nightly sessions are better predictors rather than overall use intensity.

\subsection{Difference between clusters with negative well-being}
The main difference between the Clusters 4 and 5 of people who struggle with negative emotional well-being is that negative effects are more focused onto working hours, but tend to subside or even reverse during night-time and weekend. The main difference to Cluster 4 in the phone use habits is that they tend to keep the phone in vibration mode, and that while they use it during night-time, they avoid doing email, which we consider work-related. We hypothesize that members of Cluster 5 seem to find it easier to disconnect from stressful working days. This is corroborated by the comparably high scores in emotional stability. The use of the vibration mode indicates that the stress might rather be induced by external factors (\emph{e.g.} stressful job) rather than internal factors (\emph{e.g.} emotional instability).

\subsection{Limitations}
One limitation of our study is its focus on Android phone users. Even though this results in the large majority ($\sim$90\%) of users in the country of study, there is a small percentage of users that was not included.
Moreover, this study focuses on specific emotional scales to reflect negative well-being such as PHQ-8. More scales that measures anxiety, addiction or boredom should be examined in future extension of this work.

\section{Conclusions}

On the basis of phone use logs from 340 Android users, we identified five types of typical phone use: limited use, business use, power use, and personality- \& externally induced problematic use. While previous work largely predicted that high intensity phone use predicts negative outcomes, membership of the Power Use cluster did not predict negative well-being. Instead, members of the two clusters associated with negative well-being had only average phone use. Moreover, nightly use of messaging and having the ringer mode in a setting other than normal mode is what set the two problematic-use clusters apart from the other three clusters. 
This finding indicates that instead of vilifying phone use in general, we need awareness that the \emph{type} of phone use might be a much stronger indicator than the overall phone-use intensity. This highlights the need for a more nuanced approach to studying and understanding the underlying mental problems, and not let ourselves fall into the old moral ``new technology is bad'' panic\footnote{https://www.engadget.com/2018/02/09/new-tech-addictions-are-mostly-just-old-moral-panic/}. 
Future work should attempt to replicate / falsify our findings that nightly use rather than overall use is a predictor of negative well-being. Finally, further future work should explore why the ringer mode is such a strong indicator of well-being.


%
%
%
%
%

\balance{}

\bibliographystyle{SIGCHI-Reference-Format}
\bibliography{references}

\end{document}